# INFLUENCE OF ELASTIC OSCILLATIONS ON NUCLEATION IN METALS

A.S. Nuradinov[1], O.V. Chistyakov[1], K.A. Sirenko[1], I.A. Nuradinov[1], D.O. Derecha[1],[2]*

[1]Physico-Technological Institute of Metals and Alloys of the National Academy of Sciences of Ukraine.Acad. Vernadskoho 34/1, Kyiv, Ukraine, 03680

[2]V.G. Baryakhtar Institute of Magnetism of the National Academy of Sciences of Ukraine. Acad. Vernadskoho 36b, Kyiv, Ukraine, 03142

Corresponding author e-mail: dderecha@gmail.com

**ABSTRACT**

This work is devoted to establishing the mechanisms of elastic oscillation influence on nucleation processes in metal melts. The method of physical modeling with low-temperature metallic alloys (Wood and Rose) and transparent organic media (salol, camphene, diphenylamine) was used. It was established that vibration and ultrasound significantly reduce the supercooling required to initiate crystallization. The effectiveness of the influence significantly increases for samples with solid substrates. The hypotheses about the influence through changes in melt viscosity and the exclusive role of cavitation were experimentally refuted. The transition from pre-cavitation to cavitation ultrasound regime is not accompanied by qualitative changes in the influence on nucleation. The mechanism of elastic oscillation influence is substantiated, which consists in mechanical impact on adsorbed crystal nuclei on the surfaces of solid substrates. Elastic oscillations increase the nucleation rate by creating growth steps (dislocations) on the surfaces of adsorbed nuclei as a result of mechanical friction of solid substrates and cavitation erosion. The results have fundamental significance for understanding the physical nature of metal crystallization and practical application for developing technologies for controlling structure formation.

**INTRODUCTION**

The processes of nucleation and growth of crystals in metal melts are the fundamental basis for the formation of structure and properties of cast billets. Structure control through crystallization process management remains one of the key problems of modern metallurgy, since the properties of the final product directly depend on grain size, crystal morphology, and their spatial distribution. Theoretical and experimental studies of nucleation mechanisms have a history of more than a century, but many aspects of these processes, especially under the influence of external physical factors, remain debatable. Classical nucleation theory, developed by Gibbs, Volmer, Weber, Becker, and Döring [1-3], established the thermodynamic foundations of nucleation. The thermodynamic stimulus for crystallization is the difference in Gibbs free energies between the liquid and solid phases, directly proportional to the supercooling of the melt. However, classical theory predicts significantly higher supercoolings for homogeneous nucleation than are observed in real processes, indicating the predominance of heterogeneous nucleation mechanisms [4, 5]. Heterogeneous nucleation on the surfaces of solid substrates occurs at lower supercoolings due to a reduction in the energy barrier. Modern studies have shown that adsorption of melt atoms on the substrate surface is the first stage of nucleus formation [6, 7].

Among various physical influences on melts, elastic oscillations have attracted special attention: vibration (0-1000 Hz) and ultrasound (above 16 kHz). Modern studies of ultrasonic treatment of metal melts have revealed multiple effects of structure refinement [8-10]. Experimental studies on aluminum and magnesium alloys confirmed the formation of fine-grained equiaxed structure with grain sizes of 20-200 μm depending on treatment regimes [11-13]. It was established that the most effective refinement is achieved when applying ultrasound at the nucleation stage, which indicates the crucial role of nucleation initiation [14, 15]. Mechanical vibration effectively reduces grain size, changes the morphology of primary phases, and increases casting density [16-18].

Despite the confirmed effectiveness of elastic oscillations, the physical mechanisms of their influence remain debatable. Cavitation, thermodynamic, hydrodynamic, and mechanical hypotheses have been proposed [19]. The cavitation hypothesis postulates that bubble collapse creates local hot spots, leading to temperature and density fluctuations [20]. However, effective refinement is also observed at pre-cavitation intensities, which questions the exclusive role of cavitation [21, 22]. An alternative hypothesis relates the influence to changes in viscosity and interfacial energy, but systematic viscometric measurements have not confirmed significant changes [23].

A fundamental contribution to understanding crystal growth was made by the dislocation growth theory of Burton, Cabrera, and Frank [24, 25]. The emergence of a screw dislocation on the surface creates a growth step, allowing crystals to grow at low supersaturations. Experimental observations confirmed the universality of the dislocation mechanism for a wide class of materials [26]. Applying this concept to metal solidification opens new perspectives for understanding the influence of elastic oscillations. Mechanical impact on adsorbed nuclei can create dislocations and defects on their surface, activating them for growth [27, 28].

The inconsistency of research results is largely related to the lack of reliable methods for recording the moment of crystal nucleation and differences in experimental conditions [29]. An alternative approach is the use of physical modeling with low-temperature alloys or transparent organic substances, which allows direct visual observations [30, 31].

The purpose of this study is to establish the mechanisms of elastic oscillation influence on primary crystal nucleation processes in metal melts through systematic experiments using the physical modeling method.

**Materials and Methods**

In the production of metallurgical billets, crystallization of relatively large volumes of metals usually occurs. Studying nucleation mechanisms in such objects is an extremely complex task due to the influence of numerous uncontrolled factors, namely: high temperature and opacity of metal melts, impossibility of controlling thermal and hydrodynamic processes, etc. To solve this problem, indirect research methods are widely used, particularly the physical modeling method applied in this work. The objects for research were low-temperature metallic alloys Wood (12.5% Sn, 12.5% Cd, 25% Pb, 50% Bi) and Rose (25% Sn, 25% Pb, 50% Bi) and transparent organic substances diphenylamine ($C_{12}H_{11}N$), camphene ($C_{10}H_{16}$), and salol ($C_{13}H_{10}O_3$). To study the influence of elastic oscillations on the nucleation process in melts of model alloys and media, a methodology was developed and a special experimental setup was created (Fig. 1).

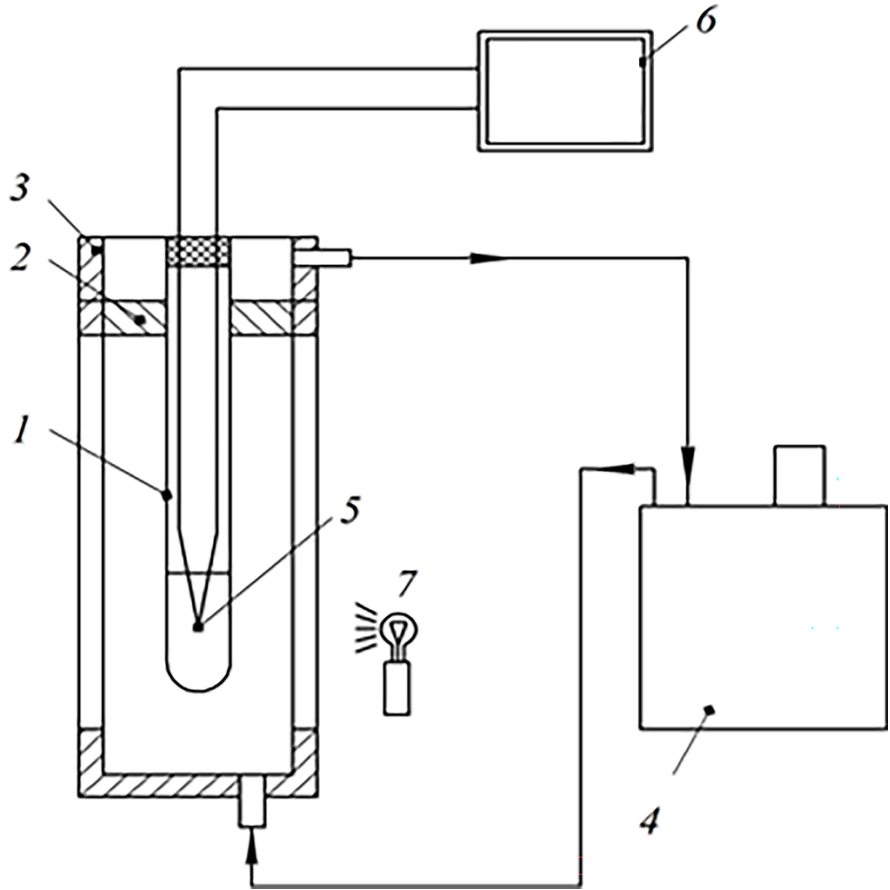

Fig. 1. Scheme of the experimental setup

The experiments were conducted according to the following methodology. First, from each model material, three test samples of equal volume were prepared in quartz tubes (Ø 8 mm) using electronic scales (with weighing accuracy ±0.01 g), and the supercooling at which spontaneous crystal nucleation occurs was determined. To ensure identical melting and crystallization conditions for model alloys, all three samples 1 were simultaneously placed in cuvette 3 using cassette 2, into which heat carrier (water) with the required temperature (sufficient for melting and superheating of the model alloy) was supplied from thermostat 4. After holding samples in the superheated state (2 min), their cooling began at a specified rate. To record the temperature of crystal formation, thermocouples 5 were installed in tubes with the test medium 1, the signal from which was displayed as absolute temperature values in digital form on the screen of digital potentiometer 6 and stored on a memory card. Using data from the memory card and a laptop with special software, temperature cooling curves of the melts of test materials were obtained. By characteristic features on these curves (i.e., when plateaus appear due to heat of crystallization release), the magnitude of melt supercooling at which crystal nucleation occurs was determined. In transparent organic media, crystal nucleation was also observed visually. To improve visual observation of crystallization processes in model media, cuvette 3 was illuminated with light from lamp 7.

Similarly, the supercooling at which crystal nucleation occurs in test samples was determined under the influence of elastic oscillations (vibration and ultrasound). To treat test samples with low-frequency elastic oscillations, an eccentric-type vibrator was attached to the lower part of the cuvette. To apply ultrasonic oscillations to the samples, the wave emitter was lowered into the water in the cuvette with tubes. The parameters of elastic oscillations were regulated within the ranges: vibration – $A = 0 \div 1$ mm, $\nu = 0 \div 100$ Hz, $P_{max} = 250$ W; ultrasound – $A = 100 \div 200$ μm, $\nu = 0 \div 38$ kHz, $I = 0 \div 5000$ W/m². In one variant of experiments, a certain amount of solid particles was introduced into tubes with test samples to simulate the presence of insoluble mechanical impurities (substrates). Quartz particles (size ~1 mm) in the amount of 10 pieces were added to samples from transparent organic media, and steel balls (Ø 1 mm) in the amount of 10 pieces were added to metallic alloys (Wood and Rose).

**Results and Discussion**

The thermodynamic stimulus for the transition of metal melt to a solid state is the difference in free energies ΔG between its liquid $G_l$ and solid $G_s$ phases. This energy difference is the driving force of the metal crystallization process and equals [32]:

$$\Delta G = L \cdot \Delta t^- / t_{cryst} \qquad (1)$$

where L is the heat of crystallization of the metal; $\Delta t^-$ is the supercooling of the metal melt; $t_{cryst}$ is the equilibrium crystallization temperature of the metal. Thus, from dependence (1), we see that for crystallization of liquid metal, its supercooling $\Delta t^-$ is necessary, the magnitude of which depends on the nature of the metal, its purity, superheating and cooling temperatures, etc. Therefore, the temperature at which nucleation begins in a metal melt cannot be taken as its thermal characteristic, similar to $t_{cryst}$. From the above, it follows that supercooling is one of the main parameters of the nucleation process in liquid metals, by the change of which one can judge the effects of impurities, external influences, and other factors.

It is known that with increasing superheat temperature of liquid metal above the liquidus temperature ($\Delta t^+$), an increase in supercooling ($\Delta t^-$) at which crystals nucleate occurs. In the literature, there are theories of surface and volume crystallization that explain the dependence of melt supercooling on superheat [31, 33, 34, 40]. In the first case, it is believed that crystallization begins on the surfaces of solid substrates (walls of casting molds, insoluble impurities, oxide films, etc.), and superheat affects the state of this surface. The second theory considers the existence in the metal melt of atomic micro-groupings (clusters), which are preserved at small superheat of the melt above the liquidus and serve as nuclei during its crystallization but are destroyed at larger superheats. But both theories do not find experimental confirmation for some metals (for example, aluminum alloys), in which maximum supercoolings are achieved at superheats greater than 200-400°C [33, 34].

From the results of our experiments, it follows that the influence of superheat on supercooling is most likely due to the state of limitedly soluble impurities in metal melts [12, 16]. This means that with increasing melt temperature, the solubility of such impurities increases, as a result of which its physicochemical properties change (equilibrium crystallization temperature, viscosity, heat of phase transition, heat capacity, etc.). In such a case, as a rule, a shift of the metastability boundary to the region of smaller supercooling occurs, which we experimentally confirmed on transparent model media (diphenylamine, salol). It was established that at superheat temperatures above a certain level, their melts noticeably changed color, i.e., irreversible processes probably occurred in them. At the same time, the melting (crystallization) temperatures of these media decreased by 2-4°C depending on their superheat temperatures [12].

Based on the above, it was assumed that external force influences (vibration, ultrasound, etc.) can be effective techniques for controlling the crystallization process of metals. Indeed, for all studied alloys and media, the influence of elastic oscillations (vibration and ultrasound) on test samples led to a decrease in the supercooling at which crystal nucleation occurred (Figs. 2, 3). At the same time, the effectiveness of elastic wave influence noticeably increased with increasing sample volumes. We see that for both media, the value of maximum supercooling ($\Delta t^-$) first sharply decreases with increasing sample volume (V). Then, upon exceeding some value of V, it practically reaches a plateau, i.e., $\Delta t^- \sim$ const. The obtained character of the dependence of $\Delta t^-$ on V is probably related to the change in the number of active impurities with increasing volume of test samples. Obviously, the larger the sample volume, the higher the probability of the presence of impurities capable of becoming crystallization centers at the minimum supercooling for it.

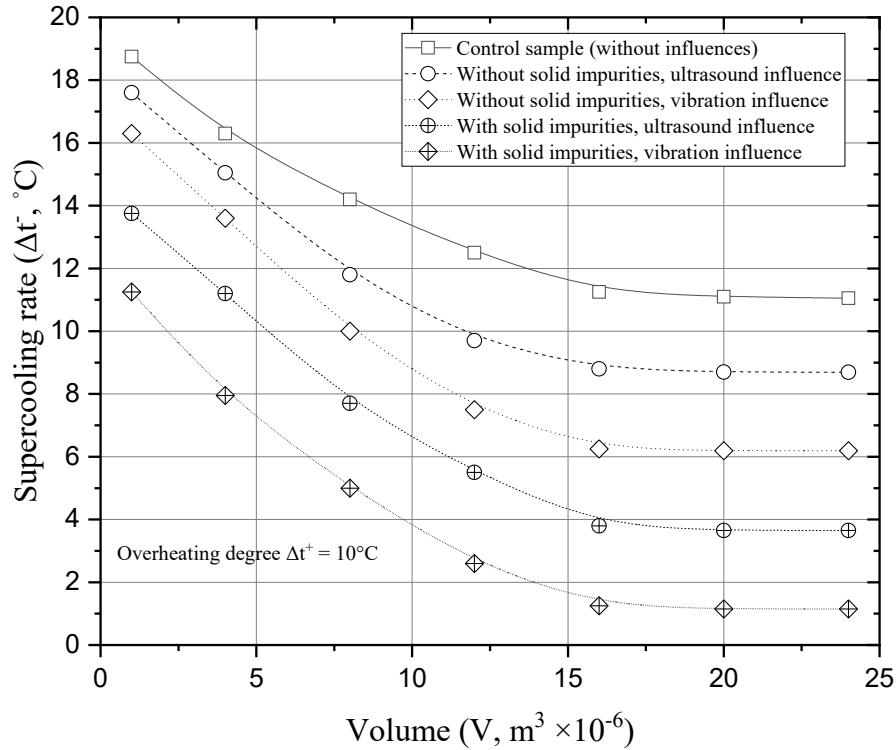

Fig. 2. Dependence of supercooling of Wood alloy on test sample volume at superheat $\Delta t^+ = 10°C$: 1 – control curve (without influences); 2, 4 – samples without solid impurities (2) and with solid impurities (4) under ultrasound influence with intensity I = 2000 W/m²; 3, 5 – samples without solid impurities (3) and with solid impurities (5) under vibration action A = 0.5 mm, ν = 50 Hz, and P = 200 W

In general, from the conducted studies, it should be noted that for all model materials, the influence of vibration on their supercooling turned out to be stronger than that of ultrasound (Figs. 2 and 3; Table 1). At the same time, the effect of the influence of elastic oscillations of both types is significantly higher for test samples into which additional solid substrates were introduced – quartz grains and steel balls (Figs. 2 and 3, curves 4 and 5; Table 1).

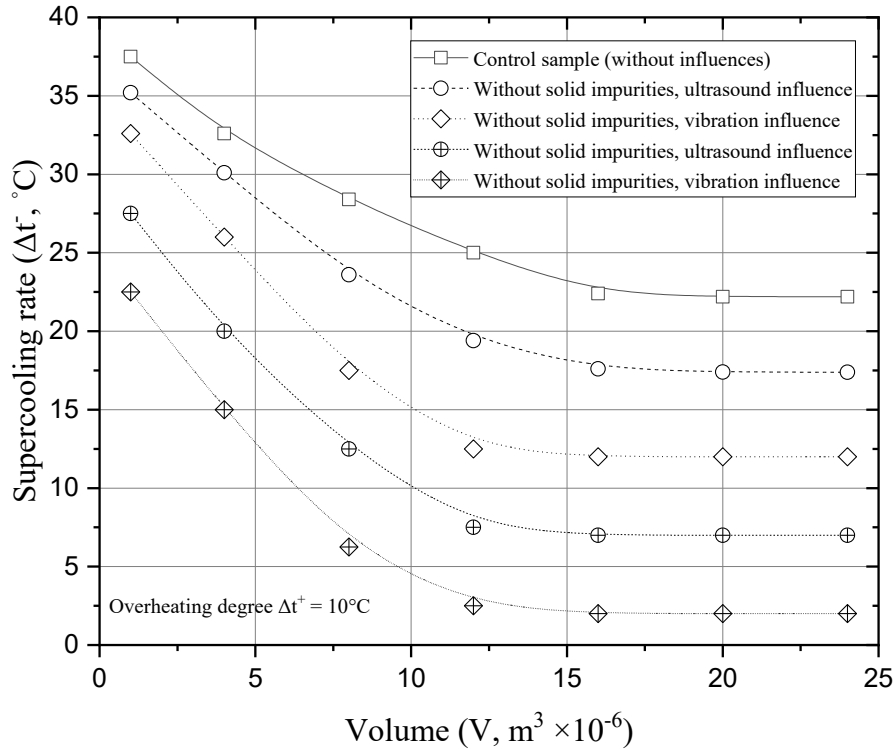

Fig. 3. Dependence of salol supercooling on test sample volume at superheat $\Delta t^+ = 10°C$: 1 – control curve (without influences); 2, 4 – samples without solid impurities (2) and with solid impurities (4) under ultrasound influence with intensity I = 2000 W/m²; 3, 5 – samples without solid impurities (3) and with solid impurities (5) under vibration action A = 0.5 mm, ν = 50 Hz, and P = 200 W

In the literature, there is no consensus regarding the mechanism of elastic wave influence on the reduction of supercooling before crystallization (i.e., on crystal nucleation) in metals. Various factors are given that play a decisive role in the nucleation process in metal melts, namely: reduction of viscosity and, accordingly, decrease in interfacial energy; appearance of temperature and density fluctuations associated with local pressure changes in cavitation, compression, and tension regions [35, 36]. From the analysis of the results of these studies, it can be stated that none of the listed factors, in our opinion, is key in reducing the supercooling of metal melts before crystallization.

Table 1. Influence of external actions on supercooling of model materials (for samples with volume V=16·10⁻⁶ m³ at their superheat $\Delta t^+ = 10°C$)

| # | Material | Supercooling, $\Delta t^-$, °C | | |
|---|---|---|---|---|
| | | Spontaneous | Under the influence of vibration $A = 0,5$ mm, $v = 50$ Hz, $P = 200$ W | Under the influence of ultrasound $I = 2000$ W/m² |
| 1. | Vood Alloy | 12/12 | 4/2 | 8/6 |
| 2. | Rose Alloy | 8/8 | 3/1 | 6/3 |
| 3. | Diphenylamine | 21/21 | 7/2 | 9/3 |
| 4. | Camphene | 14/14 | 4/2 | 6/3 |
| 5. | Salol | 22/22 | 7/2 | 17/12 |

*Note: in the numerator – value for samples without mechanical impurities; in the denominator – for samples with mechanical impurities.*

To establish the influence of elastic oscillations on the viscosity of test media, the following experiment was conducted. Viscometer with melt of test media was placed in a cuvette (Fig. 1) and subjected to the influence of elastic oscillations (vibration and ultrasound) at water temperatures close to their $t_{cryst}$. Kinematic viscosity ($\mu$) was calculated using the following formula [37]:

$$\mu = C_v \frac{g}{980,1} \cdot \tau, \qquad (2)$$

where $C_v$ is the viscometer constant; g is the acceleration of free fall; $\tau$ is the time of medium flow-down.

Fig. 4 shows the dependence of kinematic viscosity of test organic media (diphenylamine, salol, and camphene) on ultrasound intensity. We see that the influence of ultrasound of various intensities does not change the viscosity of melts of these media at temperatures close to their crystallization temperatures. Similarly to ultrasound, vibrational influence of different powers also did not affect the kinematic viscosity of model medium melts. Accordingly, it can be stated that the influence of elastic oscillations on crystal nucleation in them is not related to a change in interfacial energy, which directly depends on melt viscosity before crystallization [36].

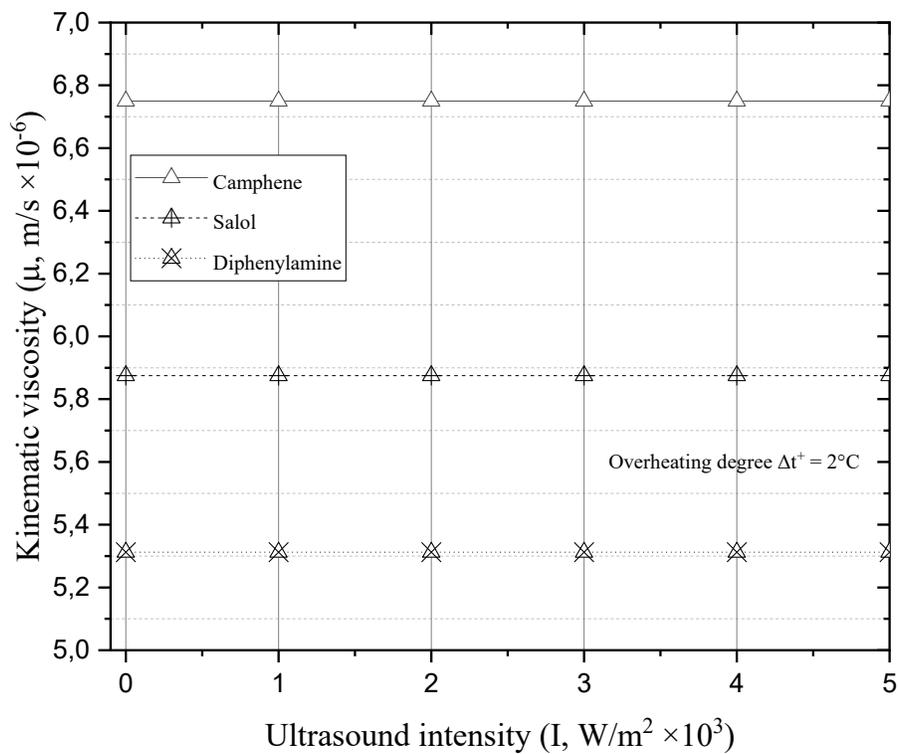

Fig. 4. Kinematic viscosity of model media depending on ultrasound intensity at $\Delta t^+ = 2°C$: 1 – camphene; 2 – salol; 3 – diphenylamine

The effectiveness of the influence of both types of elastic oscillations on the supercooling of all media also increases with increasing their powers (Figs. 5 and 6). These dependencies are not linear and have the same character of changes for all volumes. In addition, samples of larger volume are more sensitive to these influences due to, as noted above, an increase in the amount of active impurities in them. At the same time, it is important to note that the transition of ultrasonic oscillations from pre-cavitation regime to cavitation level is not manifested in any way in the character of the dependence of supercooling on their intensity (Fig. 5). The transition point of ultrasound from pre-cavitation to cavitation level in these experiments was the intensity value $I_{cav}$

= 1500 W/m² (at a frequency of 15 kHz) and was determined experimentally. At ultrasound intensity $I > I_{cav}$, no dents and holes appeared on foil lowered into water in the cuvette, which would indicate cavitation. But at $I > I_{cav}$, holes and dents appeared on the foil, indicating a transition to cavitation regime.

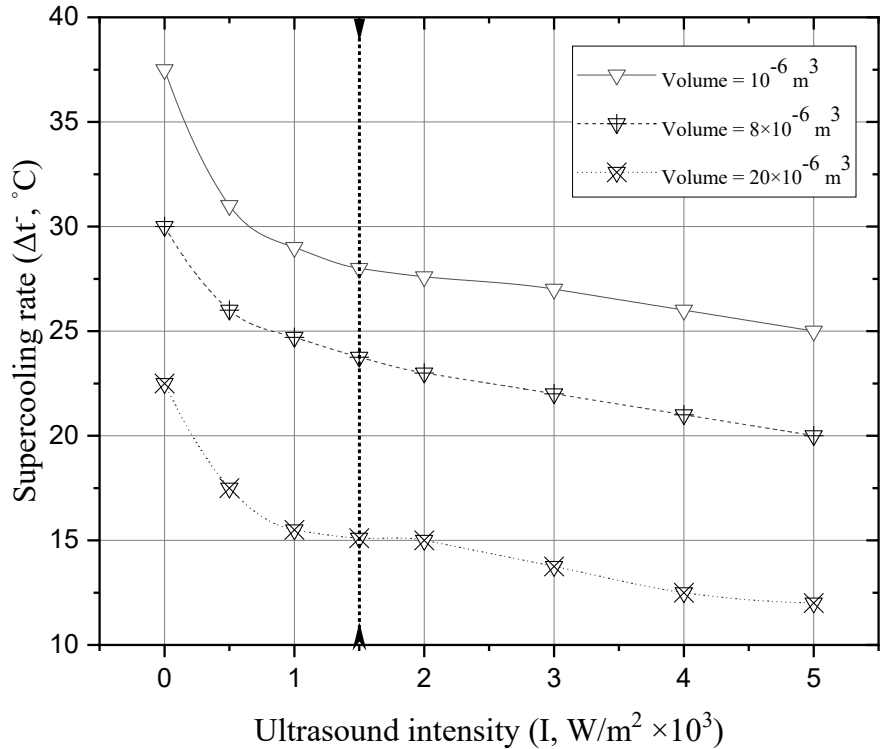

Fig. 5. Dependence of salol supercooling on ultrasound intensity:
1 – V = 10⁻⁶ m³; 2 – V = 8·10⁻⁶ m³; 3 – V = 20·10⁻⁶ m³

If the change in melting (crystallization) temperature of the model medium caused by cavitation pressure were the decisive factor in reducing its supercooling ($\Delta t^-$), then lines 1-3 in Fig. 5 in the cavitation regime region of ultrasound (i.e., to the right of the dashed line) would merge into one line. In addition, from the positions of the decisive role of thermodynamics of the cavitation process, it is impossible to explain the character of curves of the dependence of $\Delta t^-$ on I in the region of pre-cavitation ultrasound regime (i.e., to the left of the dashed line). Here, regardless of sample volumes, at ultrasound intensity when there is no cavitation, supercooling changes (decreases) faster than in the cavitation regime. With a mechanism in which cavitation would play the decisive role, everything should be the opposite.

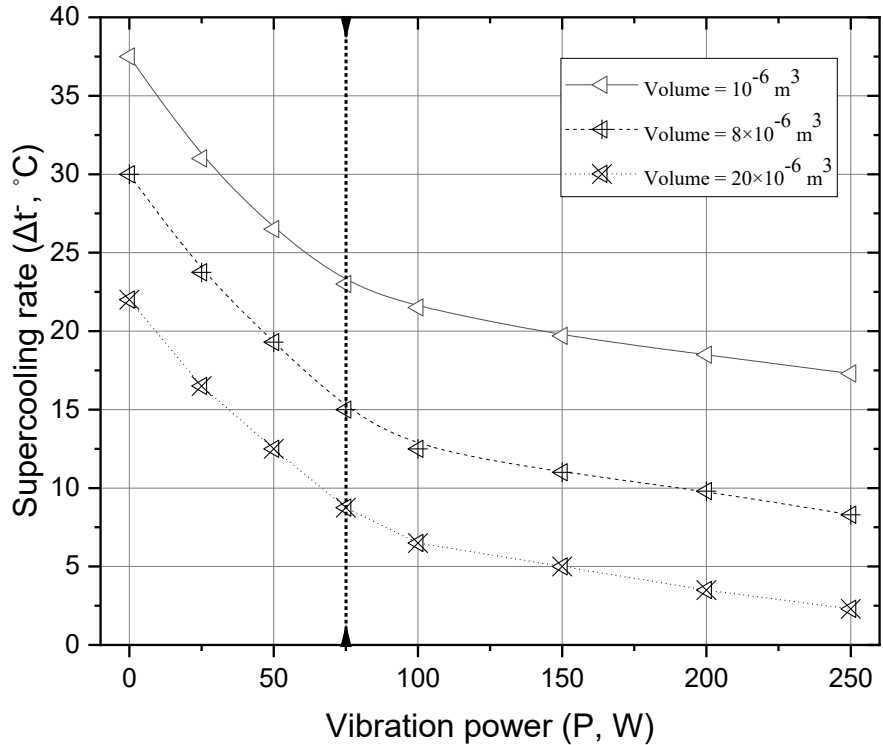

Fig. 6. Dependence of salol supercooling on vibration power:
1 – V = 10⁻⁶ m³; 2 – V = 8·10⁻⁶ m³; 3 – V = 20·10⁻⁶ m³

If we talk about local pressure changes in compression and tension regions of elastic waves, they are more than an order of magnitude smaller compared to pressures caused by cavitation [38]. Therefore, temperature and density fluctuations appearing in local compression-tension zones of metal melts under the influence of elastic waves are even less capable of playing a decisive role in crystal nucleation.

Based on the presented results, it can be assumed that the mechanism of elastic wave influence (vibration, ultrasound, etc.) on primary crystal nucleation in metal melts consists in mechanical impact on adsorbed crystal nuclei on substrate surfaces (impurities), which exist in all real alloys. Activation of solid substrates or, in other words, physical adsorption of atoms on casting mold walls, refractory insoluble particles, etc., is a generally recognized fact. Their sizes can be two to three orders of magnitude larger than theoretical values of critical nucleus radius [36, 39]. However, even such relatively large crystalline nuclei formed as a result of atom adsorption on some surface do not always grow at small supercoolings. The probability that adsorbed nuclei will become crystallization centers, in our opinion, depends on the presence of growth steps (dislocations) on their surfaces. This is also evidenced by the sharp decrease in the growth rate of salol and diphenylamine crystals in capillaries with decreasing diameter (Fig. 7) [16].

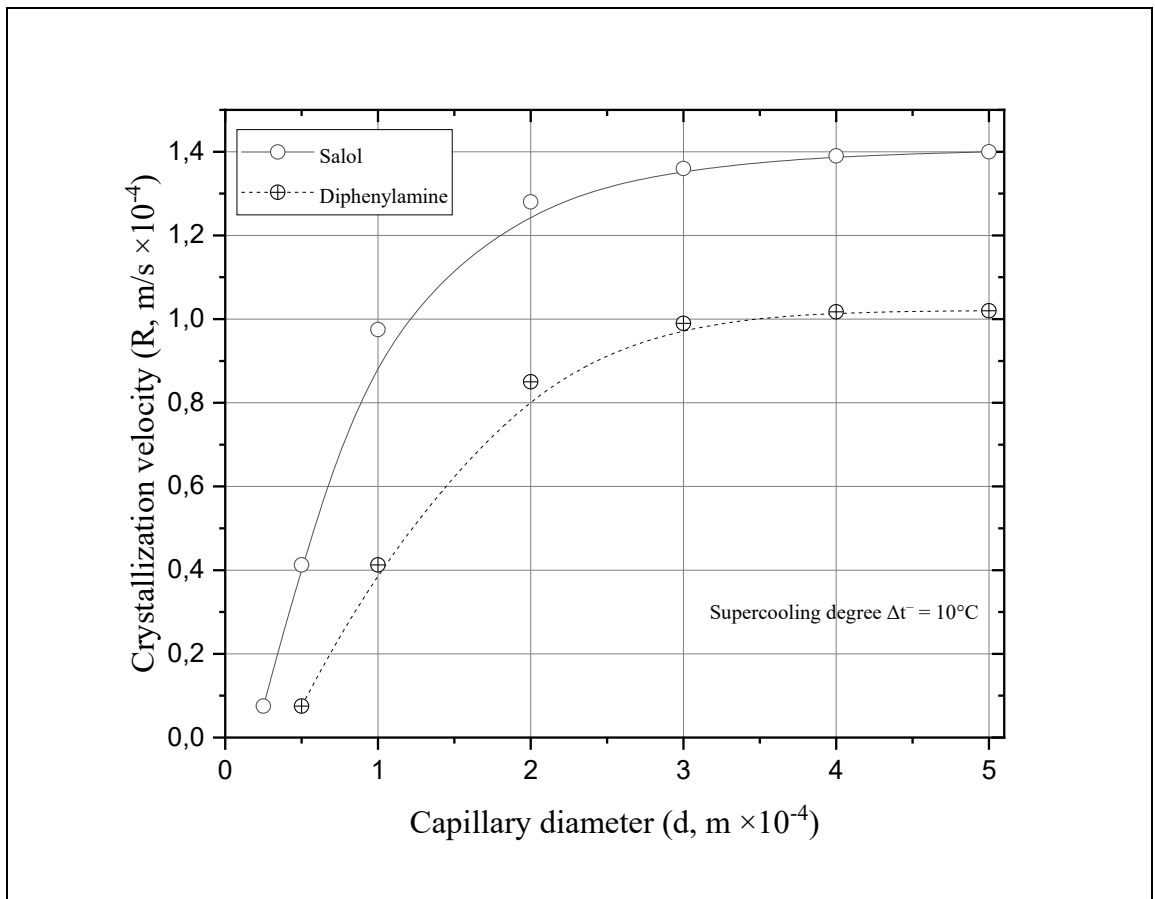

Fig. 7. Dependence of crystalline front advancement rate (R) of salol (1) and diphenylamine (2) on capillary diameter (d) at supercooling $\Delta t^- = 10°C$

When observing the growth of an individual salol crystal in capillaries of different diameters, it is possible to notice that when a crack appears at the crystal-melt boundary, a jump-like increase in its growth rate is observed. As the distance from the crack increases, the rate value decreases. Moreover, in areas where a flat crystal face advances, the rate turns out to be significantly lower than in the case of a truncated front with the formation of a toothed polycrystalline structure. Probably, the macroroughness of the interphase boundary (number of cracks, teeth, etc.) affects the number of growth steps (dislocations) and, accordingly, the rate of movement of the polycrystalline aggregate front (R). It is clear that as the capillary diameter (d) decreases at constant supercooling ($\Delta t^- $ = const), the number of protrusions on the crystallization front decreases, and with it the rate of its advancement (Fig. 7). Such a dependence of the crystallization front advancement rate (R) on capillary diameter (d) is difficult to explain otherwise than by the decisive role of growth steps (dislocations) in this process.

In the literature, crystal nucleation on substrate surfaces in metal melts is assumed to be fluctuational, i.e., caused by random detachment and attachment of wandering atoms [36, 39]. We have previously shown that nucleus formation and their growth occur not from bulk liquid but from the surface layer, which makes the dislocation mechanism more probable. At the same time, it was established that friction of solid surfaces wetted by undercooled melt is an effective method of influencing adsorbed crystal nuclei on them [16].

The influence of elastic oscillations on metal melt causes its mixing, and the maximum value of the velocity gradient ($\Delta W_m$) in this case is determined from the following equation [35]:

$$\Delta W_m = \frac{4 \cdot \pi \cdot A_{ew}}{3 \cdot \lambda},$$

where $A_{ew}$ is the amplitude of elastic waves; $\lambda$ is the wavelength.

In our opinion, with such forced mixing of metal melt, transfer of mechanical impurities contained in it occurs, and their collision with each other and with casting mold walls takes place. As a result of such collisions, growth steps (dislocations) appear on the surfaces of adsorbed nuclei through their mechanical destruction. Therefore, the presence of additional solid substrates in the form of quartz grains and steel balls in the melts of the studied model materials increases the effectiveness of both vibration and ultrasound influence (Figs. 2 and 3; Table 1). The more effective influence of vibration (compared to ultrasound) on nucleation in model alloys and media is due to different intensities of mixing of their melts (according to formula 3). In addition, an integral accompanying effect of elastic oscillation influence on metal melts is cavitation. And if cavitation occurs in immediate proximity to solid substrates, then as a result of erosive destruction of adsorbed nuclei on their surfaces, growth steps (dislocations) probably also appear for their growth.

**CONCLUSIONS**

Thus, based on the results of these studies, it was established that in melts of real metals, crystalline nuclei adsorbed on solid substrates can exist, capable of growing at significantly lower supercooling than spontaneous supercooling if mechanically acting on their surface. The influence of elastic oscillations (vibration and ultrasound) on metal melts during their cooling increases the crystal nucleation rate due to the fact that mechanical friction of solid substrates with each other and with casting mold walls, as well as cavitation erosion, create growth steps (dislocations) on the surfaces of adsorbed nuclei. Similar influence on nucleation in metal melts can have any physical influence causing their mixing. At the same time, their influence will be more effective if there are more solid substrates in metal melts capable of mechanically acting on the surfaces of adsorbed nuclei.

**REFERENCES**

[1] M. Volmer and A. Weber, "Nucleus formation in supersaturated systems," Z. Phys. Chem., vol. 119, pp. 277-301, 1926. DOI: 10.1515/zpch-1926-11927

[2] R. Becker and W. Döring, "Kinetic treatment of nucleation in supersaturated vapors," Ann. Phys., vol. 24, pp. 719-752, 1935. DOI: 10.1002/andp.19354160806

[3] D. Turnbull and J. C. Fisher, "Rate of nucleation in condensed systems," J. Chem. Phys., vol. 17, no. 1, pp. 71-73, 1949. DOI: 10.1063/1.1747055

[4] J. W. Christian, The Theory of Transformations in Metals and Alloys, 3rd ed. Oxford: Pergamon Press, 2002. DOI: 10.1016/B978-008044019-4/50022-2

[5] K. F. Kelton and A. L. Greer, Nucleation in Condensed Matter: Applications in Materials and Biology. Amsterdam: Elsevier, 2010. DOI: 10.1016/S1470-1804(09)01506-0

[6] Z. Fan, "An epitaxial model for heterogeneous nucleation on potent substrates," Metall. Mater. Trans. A, vol. 44, pp. 1409-1418, 2013. DOI: 10.1007/s11661-012-1495-8

[7] H. Men and Z. Fan, "Prenucleation induced by crystalline substrates," Metall. Mater. Trans. A, vol. 49, pp. 2766-2777, 2018. DOI: 10.1007/s11661-018-4628-x

[8] G. I. Eskin and D. G. Eskin, Ultrasonic Treatment of Light Alloy Melts, 2nd ed. Boca Raton: CRC Press, 2014. DOI: 10.1201/b17270

[9] J. Mi, D. Tan, and T. Lee, "In situ synchrotron X-ray study of ultrasound cavitation and its effect on solidification microstructures," Metall. Mater. Trans. B, vol. 46, pp. 1615-1619, 2015. DOI: 10.1007/s11663-014-0256-z

[10] D. G. Eskin, J. Mi, and M. Khavari, "Application of a pulsed magnetic field during solidification of direct chill cast aluminum alloys: Part 2. Microstructure evolution," Metall. Mater. Trans. A, vol. 53, pp. 2767-2784, 2022. DOI: 10.1007/s11661-022-06710-x

[11] T. V. Atamanenko, D. G. Eskin, L. Zhang, and L. Katgerman, "Criteria of grain refinement induced by ultrasonic melt treatment of aluminum alloys containing Zr and Ti," Metall. Mater. Trans. A, vol. 41, pp. 2056-2066, 2010. DOI: 10.1007/s11661-010-0232-4

[12] H. T. Li, Y. Wang, and Z. Fan, "Mechanisms of enhanced heterogeneous nucleation during solidification in binary Al-Mg alloys," Acta Mater., vol. 60, no. 4, pp. 1528-1537, 2012. DOI: 10.1016/j.actamat.2011.11.044

[13] A. Ramirez and M. Qian, "Potency of high-intensity ultrasonic treatment for grain refinement of magnesium alloys," Scripta Mater., vol. 59, no. 1, pp. 19-22, 2008. DOI: 10.1016/j.scriptamat.2008.02.017

[14] G. I. Eskin, "Principles of ultrasonic treatment: Application for light alloys melts," Adv. Performance Mater., vol. 4, pp. 223-232, 1997. DOI: 10.1023/A:1008603815525

[15] D. G. Eskin, "Broad prospects for commercial application of the ultrasonic (cavitation) melt treatment of light alloys," Ultrason. Sonochem., vol. 8, no. 3, pp. 319-325, 2001. DOI: 10.1016/S1350-4177(00)00074-2

[16] M. Haghayeghi, H. Bahai, and P. Kapranos, "Effect of ultrasonic argon degassing on dissolved hydrogen in aluminium alloy," Mater. Lett., vol. 82, pp. 230-232, 2012. DOI: 10.1016/j.matlet.2012.05.087

[17] O. V. Abramov, "Ultrasound in Liquid and Solid Metals," CRC Press, 1994, ISBN: 978-0849391903.

[18] J. Zhang, Z. Fan, Y. Q. Wang, and B. L. Zhou, "Effect of electromagnetic vibration on microstructure and properties of rapidly solidified Mg-Zn-Y-Zr alloys," Mater. Sci. Eng. A, vol. 448, nos. 1-2, pp. 189-194, 2007. DOI: 10.1016/j.msea.2006.10.048

[19] I. Campbell, "Effects of vibration during solidification," Int. Metals Rev., vol. 26, no. 2, pp. 71-108, 1981. DOI: 10.1179/imr.1981.26.1.71

[20] K. S. Suslick and G. J. Price, "Applications of ultrasound to materials chemistry," Annu. Rev. Mater. Sci., vol. 29, pp. 295-326, 1999. DOI: 10.1146/annurev.matsci.29.1.295

[21] A. Eldarchanov, O. Figovsky, and A. Nuradinov, "Influence of vibration on continuously cast blooms," Scientific Israel - Technological Advantages, vol. 16, nos. 1-2, pp. 225-228, 2014.

[22] N. S. Barekar, S. Tzamtzis, N. H. Babu, J. Patel, H. V. Atkinson, and Z. Fan, "Processing of aluminum-graphite particulate metal matrix composites by advanced shear technology," J. Mater. Eng. Perform., vol. 18, pp. 1230-1240, 2009. DOI: 10.1007/s11665-009-9372-8

[23] G. I. Eskin, D. G. Eskin, and M. Taghavi, "Influence of ultrasonic melt treatment on the formation of primary intermetallics and related grain refinement in aluminum alloys," J. Mater. Sci., vol. 46, pp. 5528-5538, 2011. DOI: 10.1007/s10853-011-5499-3

[24] W. K. Burton, N. Cabrera, and F. C. Frank, "The growth of crystals and the equilibrium structure of their surfaces," Phil. Trans. R. Soc. Lond. A, vol. 243, no. 866, pp. 299-358, 1951. DOI: 10.1098/rsta.1951.0006

[25] F. C. Frank, "The influence of dislocations on crystal growth," Discuss. Faraday Soc., vol. 5, pp. 48-54, 1949. DOI: 10.1039/DF9490500048

[26] P. G. Vekilov, "Nucleation," Cryst. Growth Des., vol. 10, no. 12, pp. 5007-5019, 2010. DOI: 10.1021/cg1011633

[27] S. Kumar, C. Davi, P. R. Murthy, A. P. Moon, and G. Phanikumar, "Effect of ultrasonic shot peening on microstructure and mechanical properties of 7075-T651 aluminum alloy," J. Mater. Eng. Perform., vol. 28, pp. 4288-4298, 2019. DOI: 10.1007/s11665-019-04192-x

[28] Y.-J. Liu, C.-L. Peng, W.-M. Li, X.-L. Zhu, M.-G. Shen, X.-W. Liao, K. Liu, C.-Y. Wei, Y. A. Yusuf, J. Yang, and C.-Y. Cai, "Effect of hollow insulation riser on shrinkage porosity and solidification structure of ingot," J. Iron Steel Res. Int., vol. 29, no. 12, pp. 1951-1960, 2022. DOI: 10.1007/s42243-022-00790-8

[29] Z. Fan and Y. Wang, "Impeding nucleation for more significant grain refinement," Sci. Rep., vol. 10, art. 9448, 2020. DOI: 10.1038/s41598-020-66190-8

[30] T. Yuan, S. Kou, and Z. Luo, "Grain refining by ultrasonic stirring of the weld pool," Acta Mater., vol. 106, pp. 144-154, 2016. DOI: 10.1016/j.actamat.2016.01.016

[31] A. S. Nuradinov, A. V. Nogovitsyn, K. A. Sirenko, I. A. Nuradinov, et al., "Physical simulation of nucleation and crystallization processes in transparent organic melts," Phys. Scr., vol. 100, art. 035909, 2025. DOI: 10.1088/1402-4896/adadb0

[32] M. C. Flemings, Solidification Processing. New York: McGraw-Hill, 1974.

[33] D. A. Porter and K. E. Easterling, Phase Transformations in Metals and Alloys, 2nd ed. London: Chapman & Hall, 1992. DOI: 10.1007/978-1-4899-3051-4

[34] J. W. Christian, The Theory of Transformations in Metals and Alloys, Part I, 2nd ed. Oxford: Pergamon Press, 1975. DOI: 10.1016/B978-0-08-044019-4.50022-2

[35] G. I. Eskin and D. G. Eskin, "Production of natural and synthesized aluminum-based composite materials with the aid of ultrasonic (cavitation) treatment of the melt," Ultrason. Sonochem., vol. 10, no. 4-5, pp. 297-301, 2003. DOI: 10.1016/S1350-4177(02)00158-X

[36] W. Kurz and D. J. Fisher, Fundamentals of Solidification, 4th ed. Zurich: Trans Tech Publications, 1998. ISBN: 978-0878494697

[37] ASTM D445-19a, "Standard Test Method for Kinematic Viscosity of Transparent and Opaque Liquids," ASTM International, West Conshohocken, PA, 2019. DOI: 10.1520/D0445-19A

[38] L. D. Rozenberg, Ed., High-Intensity Ultrasonic Fields. New York: Plenum Press, 1971. DOI: 10.1007/978-1-4757-5408-7

[39] A. L. Greer, A. M. Bunn, A. Tronche, P. V. Evans, and D. J. Bristow, "Modelling of inoculation of metallic melts: application to grain refinement of aluminium by Al-Ti-B," Acta Mater., vol. 48, pp. 2823-2835, 2000. DOI: 10.1016/S1359-6454(00)00094-X

[40] A. S. Nuradinov, K. A. Sirenko, I. A. Nuradinov, O. V. Chystiakov, and D. O. Derecha, "The size effect on nucleation process during solidification of metals," Jul. 07, 2025, arXiv: arXiv:2507.04737. DOI: 10.48550/arXiv.2507.04737.